\documentclass[12pt]{article}
\usepackage{amsmath, amssymb, amsthm, float, fullpage, graphicx, multirow,parskip, subcaption, setspace}
\usepackage{comment}
\usepackage{tikz}
\usetikzlibrary{shapes.geometric, positioning}
\usetikzlibrary{quotes, angles}
\usepackage{rotating}
\usepackage{hyperref}
\usepackage{natbib}
\usepackage[T1]{fontenc}
\usepackage[utf8]{inputenc}
\usepackage{authblk}

\definecolor{PennRed}{RGB}{152, 30 50}
\definecolor{PennBlue}{RGB}{0, 44, 119}
\definecolor{PennGreen}{RGB}{94, 179,70}
\definecolor{PennViolet}{RGB}{141, 76, 145}
\definecolor{PennSkyBlue}{RGB}{14, 118, 188}
\definecolor{PennOrange}{RGB}{243, 117, 58}
\definecolor{PennBrightRed}{RGB}{223,82, 78}

\hypersetup{pdfborder = {0 0 0.5 [3 3]}, colorlinks = true, linkcolor = PennBrightRed, citecolor = PennSkyBlue}

\let\origthanks\thanks
\renewcommand\thanks[1]{\begingroup\let\rlap\relax\origthanks{#1}\endgroup}

\title{Expected Hypothetical Completion Probability}

\author{Sameer K. Deshpande\thanks{CSAIL MIT, email: sameerd@alum.mit.edu}  and Katherine Evans\thanks{Toronto Raptors, email: causalkathy@gmail.com}}

\begin{document}
\maketitle

\def\C{\mathbb{C}}
\def\R{\mathbb{R}}
\def\Q{\mathbb{Q}}
\def\Z{\mathbb{Z}}
\def\N{\text{N}}
\def\P{\mathbb{P}}
\def\E{\mathbb{E}}
\def\by{\mathbf{y}}
\def\bx{\mathbf{x}}
\def\bz{\mathbf{z}}
\def\bw{\mathbf{w}}

\def\bY{\mathbf{Y}}
\def\bX{\mathbf{X}}

\begin{abstract}

\singlespacing
Using high-resolution player tracking data made available by the National Football League (NFL) for their 2019 Big Data Bowl competition, we introduce the Expected Hypothetical Completion Probability (EHCP), a objective framework for evaluating plays.
At the heart of EHCP is the question ``on a given passing play, did the quarterback throw the pass to the receiver who was most likely to catch it?''
To answer this question, we first built a Bayesian non-parametric catch probability model that automatically accounts for complex interactions between inputs like the receiver's speed and distances to the ball and nearest defender.
While building such a model is, in principle, straightforward, using it to reason about a hypothetical pass is challenging because many of the model inputs corresponding to a hypothetical are necessarily unobserved.
To wit, it is impossible to observe how close an un-targeted receiver would be to his nearest defender had the pass been thrown to him instead of the receiver who was actually targeted.
To overcome this fundamental difficulty, we propose imputing the unobservable inputs and averaging our model predictions across these imputations to derive EHCP. 
In this way, EHCP can track how the completion probability evolves for each receiver over the course of a play in a way that accounts for the uncertainty about missing inputs.

\end{abstract}

\newpage
\onehalfspacing
\section{Introduction}
\label{sec:introduction}

Consider two passing plays during the game between the Los Angeles Rams and visiting Indianapolis Colts in the first week of the 2017 National Football League (NFL) season. 
The first passing play was a short pass in the first quarter from Colts quarterback Scott Tolzien intended for  T.Y. Hilton which was intercepted by Trumaine Johnson and returned for a Rams touchdown. 
The second passing play was a long pass from Rams quarterback Jared Goff to Cooper Kupp, resulting in a Rams touchdown. 
In this work, we consider the question: which play had the better route(s)? 

From one perspective, one could argue that Kupp's route was better than Hilton's; after all it resulted in the offense scoring while the first play resulted in a turnover and a defensive score.
However, ``resulting'', or evaluating a decision based only on its outcome is not always appropriate or productive. 
Two recent examples come to mind: Pete Carroll's decision to pass the ball from the 1 yard line in Super Bowl XLIX and the ``Philly Special'' in Super Bowl LII. 
Had the results of these two plays been reversed, Pete Carroll might have been celebrated and Doug Pederson criticized.

If evaluating plays solely by their outcomes is inadequate, on what basis should we compare routes? 
One very attractive option is to use tracking data. 
In the NBA, tracking data has been available since 2013, allowing researchers to quantify actions with spatial statistics. 
For instance, the XY Research group has produced several papers outlining how to organize and evaluate possessions and player ability using tracking data.
\citet{Cervone1} and \citet{Cervone2} introduced expected possession value (EPV), a framework for using player tracking data to estimate the expected number of points scored by the end of an offensive possession.
\citet{Franks1} used tracking data to quantify player ability and \citet{MillerBornn} introduced Possessions Sketches, a machine learning method that decomposes player movement into a small number of interpretable actions. 
Outside of professional basketball, there has been comparatively little work done using tracking data in the public domain.
Recently, and perhaps most similar to the method presented below, is \citet{Burke}, who develops a deep learning method to quantify quarterback decision making in the NFL.

In this paper, we will focus on completion probability and will discuss how one might adapt the methodology developed in the sequel to other measures of play success in Section~\ref{sec:discussion}.
Intuitively, we might tend to prefer routes which maximize the receiver's chance of catching the pass. 
To this end, if we let $y$ be a binary indicator of whether a pass was caught and let $\bx$ be a collection of covariates summarizing information about the pass, we can consider a logistic regression model of completion probability:
\begin{equation}
\label{eq:completion_probability_model}
\log{\left(\frac{\P(y = 1 | \bx)}{\P(y = 0 | \bx)}\right)} = f(\bx),
\end{equation}
or equivalently $\P(y = 1 | \bx) = \left[1 + \text{e}^{-f(\bx)}\right]^{-1},$ for some unknown function $f.$

If we knew the true function $f,$ assessing a route is easy: we simply plug in the relevant covariates $\bx$ and compute the forecasted completion probability.
Regardless of whether the receiver caught the actual pass, if this forecasted probability exceeded 50\%, we could conclude that the route was run and the pass was thrown in a way that gave the receiver a better chance than not of catching the pass.  
We could moreover directly compare the forecasted completion probability of the two plays mentioned above; if it turned out that the Tolzien interception had a higher completion probability than the Kupp touchdown, that play would not seem as bad, despite the much worse outcome. 
While such a comparison is a good first step towards evaluating routes, it is not completely satisfactory -- there are often multiple receivers running routes on a play and this comparison focuses only on a single player's chances of successfully catching a specific pass thrown to a single location along his route.
The comparison does not, in particular, answer the very natural follow-up question: was there another location along a possible \emph{different} receiver's route where the completion probability was higher? 
If so, one could argue that the quarterback ought to have thrown to ball to that spot. 

\begin{figure}[H]
\includegraphics[width = 1.0\textwidth]{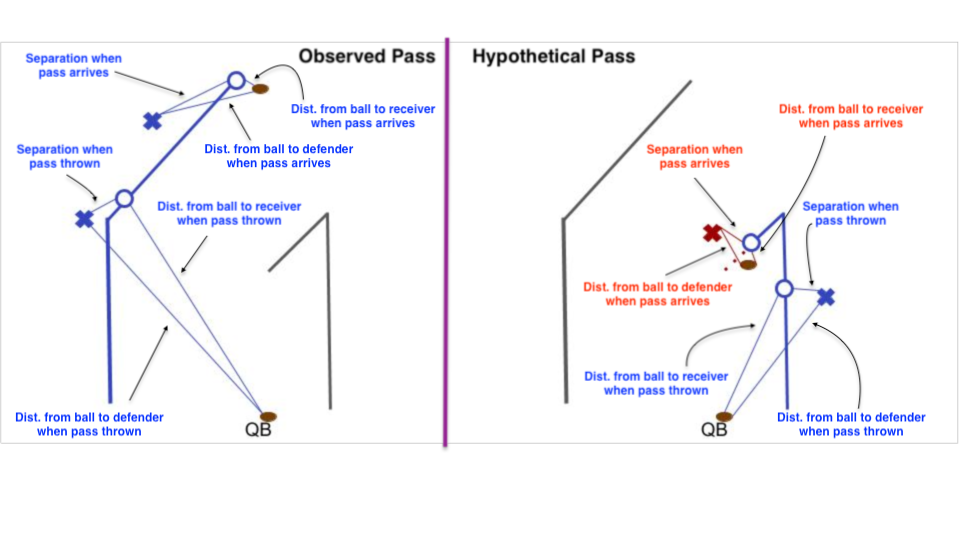}
\caption{Schematic of what we directly observe on an actual pass (left panel) from our dataset and what we cannot observe for a hypothetical pass (right panel). In both passes, there are two receivers running routes.The targeted receiver is denoted with a circle and the defender closest to the receiver is denoted with an X. Unobservables are colored red while observables are colored blue. For the hypothetical pass, we are unable to measure the pairwise distances between the targeted receiver, his closest defender, and the ball when the pass arrives. Intuitively, all of these factors are very predictive of completion probability.}
\label{fig:visual_route_manuscript}
\end{figure}

At first glance, determining the completion probability at an arbitrary location along a different receiver's route seems impossible: even if we know the true function $f,$ we are essentially trying to deduce what might have happened in a \emph{counterfactual} world where the quarterback had thrown the ball to a different player at a different time, with the defense reacting differently.
For the attempted passes that we actually observe, we are able to directly measure all possible information about the pass including, for instance, the receiver's speed and separation from the nearest defender at (i) the time that the pass was thrown and (ii) the time that the pass arrives and he attempts to catch the ball.
In contrast, on a counterfactual pass, we can only potentially observe this information up to the the time the counterfactual pass is thrown.
The fundamental challenge is that we cannot observe any covariates measured at the time the counterfactual pass arrives; see Figure~\ref{fig:visual_route_manuscript}.

Before proceeding, we pause for a moment to distinguish between our use of the term ``counterfactual'' and its use in causal inference.
The general causal framework of counterfactuals supposes that we change some treatment or exposure variable and asks what happens to downstream outcomes.
In contrast, in this work, we considering changing a midstream variable, the location of the intended receiver when the ball arrives, and then impute both upstream and downstream variables like the time of the pass and the receiver separation at the time the ball arrives. 
In this work, we use ``counterfactual'' interchangeably with ``hypothetical'' and hope our more liberal usage is not a source of further confusion below\footnote{Author's note: We use the word ``counterfactual'' interchangeably with ``hypothetical'' because while an unobserved pass is hypothetical, the intended receiver of that pass is not.}.

The difficulty in determining counterfactual completion probabilities is compounded by the fact that we do not know the true regression function $f$ and must, therefore, estimate it from observational data.
In the process, estimation uncertainty about $f$ propagates to the uncertainty about the hypothetical completion probabilities. 
We argue that an objective assessment of routes based on a completion probability must address the inherent uncertainty in the hypothetical inputs as well as uncertainty stemming from estimating the completion probability model.

In this work, we aim to overcome these challenges. Using tracking, play and game data from the first 6 weeks of the 2017 NFL season, we developed such an assessment, which we call Expected Hypothetical Completion Probability (EHCP).
At a high-level, our framework consists of two steps. 
First, we estimate the log-odds of a catch as a function of several characteristics of each observed pass in our data.
Then, we simulate the characteristics of the hypothetical pass that we do not directly observe and compute the average completion probability of the hypothetical pass.
The rest of this paper is organized as follows.
In Section~\ref{sec:ehcp_framework}, we outline the EHCP framework and describe the data used to develop it.
We describe our Bayesian procedure for fitting the catch probability model in Equation~\eqref{eq:completion_probability_model} in Section~\ref{sec:completion_probability}.
Section~\ref{sec:illustration} illustrates the EHCP framework on several routes for the both the Tolzein interception and Kupp touchdown mentioned above.
We conclude with a discussion of potential methodological improvements and refinements and potential uses of our EHCP framework.

\section{The EHCP Framework}
\label{sec:ehcp_framework}

The EHCP framework consists of three parts (i) a completion probability model, which is trained using the observational data provided by the NFL (described in Section~\ref{sec:data}), that takes as input observed features of passes and returns the estimated completion probability, (ii) an \emph{imputation} method for predicting the variables that are unobservable for hypothetical passes, and (iii) a strategy for combining these two parts and propagating uncertainty in a coherent fashion. 
In this section, we describe these three parts and also the data used. 

\subsection{The NFL Big Data Bowl Dataset}
\label{sec:data}

The NFL Big Data Bowl Dataset contains tracking, play, and game data from all 91 games in the the first 6 weeks of the 2017 NFL season. The tracking data is at the granularity of every tenth of a second per play. For each player on the field (and the ball) for a given play, the data contains: time stamp of play (time, yyyy-mm-dd, hh:mm:ss), player position along the long axis of the field (0 - 120 yards), player position along the short axis of the field (0 - 53.3 yards), speed in yards/second, distance traveled from prior time point (in yards), angle of player motion (0 - 360 degrees), tagged play details, (including moment of ball snap, pass release, pass catch, tackle, etc), player identification number (unique across players), player name, jersey number of player, team (away or home) of corresponding player, frame identifier for each play (starting at 1), unique game identifier, and play identifier (not unique across games). From the tracking data we are able to calculate eadch player's distance to the ball and to other players as well as their cumulative distance run in the play and in the game.
The play data contains (not an exhaustive list): game quarter, time on game clock at the start of the play (counting down from 15:00, MM:SS), down, distance needed for a first down, yard line at line-of-scrimmage, home team score prior to the play, visiting team score prior to the play, home team points at the end of the play, visiting team points at the end of the play, indicator for penalty called on play, indicator for special teams play, pass length (in yards), result of pass play (caught, incomplete, intercepted, run, sack), result of play in yards, and a description of the play. The play data allows us to calculate the difference in score and incorporate the timing of the play in our models.
The game data contains game specific information like final score, temperature, humidity, and wind. We did not use any of this information in our models.
More detailed information on the data can be found at the Big Data Bowl GitHub page\footnote{https://github.com/nfl-football-ops/Big-Data-Bowl}.

\subsection{Estimating Completion Probability}
\label{sec:completion_probability}

The first step of the EHCP framework is to estimate completion probability.
In order to do so, for each passing play in the BDB dataset, we extract or compute several covariates which we think are predictive of completion probability.
These covariates can broadly be divided into three categories: those we can observe at the time the pass is thrown, those that are observed when the pass arrives and the receiver attempts to catch the ball, and situation variables describing the context in which the pass was thrown.
We include the following variables measured at the time the pass was thrown into our model: (i) the receiver's speed and direction, (ii) the pairwise Euclidean, horizontal, and vertical distances between the receiver, his nearest defender, and the ball, and (iii) the total Euclidean distance the receiver has run up to that point in the game.
We also include measurements of these same variables at the time when the receiver attempts to catch the ball as well as the changes in the receiver's speed, separation, direction, and total distance travelled while the ball is in the air.
Finally, we include the time between the snap and the pass, the amount of time that the ball is in the air, the total time from snap to catch attempt, the number of seconds left in the half, down, distance, yards to go to reach a first down, whether the offensive team is leading, and a categorical variable summarizing by how many scores the offensive team is leading or trailing (9+ points, 1 -- 8 points, 0 points) \footnote{This discretization of the score differential was suggested by Mike Lopez \url{https://twitter.com/StatsbyLopez/status/1082287615485886464}}.
For each of the N = 4,913 passes in our dataset, we let $y_{i}$ be a binary indicator of whether the pass was caught and we let $\bx_{i}$ be a vector concatenating all of the $p = $ covariates for that pass.
Perhaps the simplest completion probability model we can build is a convention logistic regression:
$$
\log{\left(\frac{\P(y_{i} = 1 \mid \bx_{i})}{\P(y_{i} = 0 \mid \bx_{i})}\right)} = \bx_{i}^{\top}\theta
$$
where $\theta \in \R^{p}$ is some unknown vector of each covariates effect.
We take a Bayesian approach to estimating $\theta$ by specifying a prior $\pi(\theta),$ that captures all of our initial uncertainty about each covariates effect, and  we update it to form the posterior distribution $\pi(\theta \mid \by)$ using Bayes' theorem: $\pi(\theta \mid \by) \propto \pi(\theta)p(\by \mid \theta)$ where $p(\by \mid \theta)$ is the \textit{likelihood} implied by the logistic model in Equation~\eqref{eq:completion_probability_model}. 
Since the posterior distribution is not analytically tractable, we use a Markov Chain Monte Carlo (MCMC) simulation to generate draws $\theta^{(1)}, \ldots, \theta^{(N)}$ from the posterior.
Specifically, upon re-scaling all of the continuous covariate to have mean zero and standard deviation 0.5 and re-centering all binary covariates to have mean zero, as recommended by \citet{Gelman2008}, we place independent $N(0,1)$ priors on each element of $\theta.$
We fit this model in Stan \citep{Carpenter2017} through the interface provided by the ``rstan'' \texttt{R} package \citep{rstan}.

While straightforward to fit, a major drawback to this simple Bayesian logistic regression model is its assumption that none of the covariates interact with each other in ways that meaningfully impact the log-odds of completing a pass.
On its face, this assumption is tenuous at best, motivating us to consider estimating the unknown log-odds function $f$ with regression trees, which naturally incorporate interactions by design. 
Specifically, we use \citet{Chipman2010}'s Bayesian Additive Regression Trees (BART) to express $f$ as a sum of several regression trees.
Since it's introduction, BART has been used across a wide variety of domains and has demonstrated excellent predictive performance. 

Similar to the simple linear-logistic model above, to use BART we start by specifying a prior $\pi(f)$ mean to reflect all of our initial uncertainty about the unknown function $f.$
In the case of BART, rather than specifying this prior directly, we instead specify a prior over the space of regression trees used to approximate $f..$
We then update this prior to compute a posterior over the space of regression trees, which induces a posterior over $f.$
For a review of Bayesian tree-based methods, please see \citet{Linero2017} for further details about the BART prior and Gibbs sampler, please see \citet{Chipman2010}.
We fit the BART model using the \texttt{lbart()} function available in the ``BART'' \texttt{R} package \citep{BART}.
In order to facilitate variable selection, we actually used a slightly modified BART prior due to \citet{Linero2018}. 
Operationally, this was done by running \texttt{lbart()} with the option \texttt{sparse = TRUE}.
Our code is available at \url{https://www.github.com/skdeshpande91/ehcp}

For the purposes of constructing EHCP, we must prioritize accurate predictions of the completion probability. 
To pick between the the simple Bayesian logistic regression model and the logistic BART model, we first run a validation experiment in which we generate 10  75\%/25\% training/testing splits of our data.
For each training dataset, we fit both models and then for each pass $i$ in the testing dataset, we compute the posterior predictive mean completion probability $\hat{p}_{i}$ for each method.
To assess the predictive performance, we computed for each pass in the testing set (i) the squared error $(y_{i} - \hat{p}_{i})^{2},$ (ii) the log-loss $-(y_{i}\log{\hat{p}_{i}} + (1 - y_{i})\log{(1 - \hat{p}_{i})}),$ and the mis-classification error $\mathbf{1}(y_{i} \neq \mathbf{1}(\hat{p}_{i} \geq 0.5)).$ 
Table~\ref{tab:validation} shows the mean square error, log-loss, and mis-classification rate averaged over each training/testing split.
\begin{table}[H]
\centering
\caption{Predictive performance of the Bayesian logistic regression model and the BART-based model averaged across 10 training/testing splits of the data. Standard deviations are shown in parentheses}
\label{tab:validation}
\begin{tabular}{lrrr}
\hline
~ & Mean Square Error & Misclassification & Log-Loss \\ \hline
Bayesian Logistic & 0.099 (0.004) & 0.138 (0.008) & 0.332 (0.018) \\
BART & 0.086 (0.004) & 0.113 (0.005) & 0.289 (0.011)  \\ \hline
\end{tabular}
\end{table}

Across each performance measure, we see that the BART-based model has better predictive performance than the simpler Bayesian logistic regression model.
For that reason, we will use it in our construction of EHCP. 
We note, however, that despite its superior performance, our BART-based completion probability model is much more opaque than the simpler model.
In particular, as with any tree-based procedure, it is not immediately clear which variables are the most predictive of catch probability.
In the context of EHCP, understanding variable importance is critical: if none of the variables measured when the pass arrives at the receiver's location impacted the completion probability, we could arguably avoid the imputation of missing covariates altogether.

\citet{Linero2018} introduced a modification of BART that allows for variable selection, which we used to fit our model.
Whereas \citet{Chipman2010}'s original decision tree prior uniformly sampled the variable on which to split at each internal decision node, \citet{Linero2018} gives each variable its own ``splitting probability'' and places a Dirichlet prior over the collection of splitting probabilities.
As a result, when we fit our model with this modified prior, in addition to approximate samples from the posterior predictive completion probabilities, we also obtain draws from the posterior distribution of each variables splitting probability.
It turns out that the variables with the largest posterior mean splitting probabilities were the receiver's speed when the catch was attempted (20.74\%), the Euclidean distance between the receiver and the ball when the catch was attempted (17.29\%), the total Euclidean distance the receiver travelled between the snap and the catch attempt (10.92\%),  the Euclidean distance between the receiver and the ball when the pass was thrown (7.35\%), and the separation at the time the catch was attempted (6.74\%).
In other words, if we were to draw a decision tree from the posterior distribution, we would split observations along the receiver's speed while attempting to catch the ball just over 20\% of the time but would split observations based on the receiver's separation just under 7\% of the time.

\begin{figure}[H]
\centering
\includegraphics[width = \textwidth]{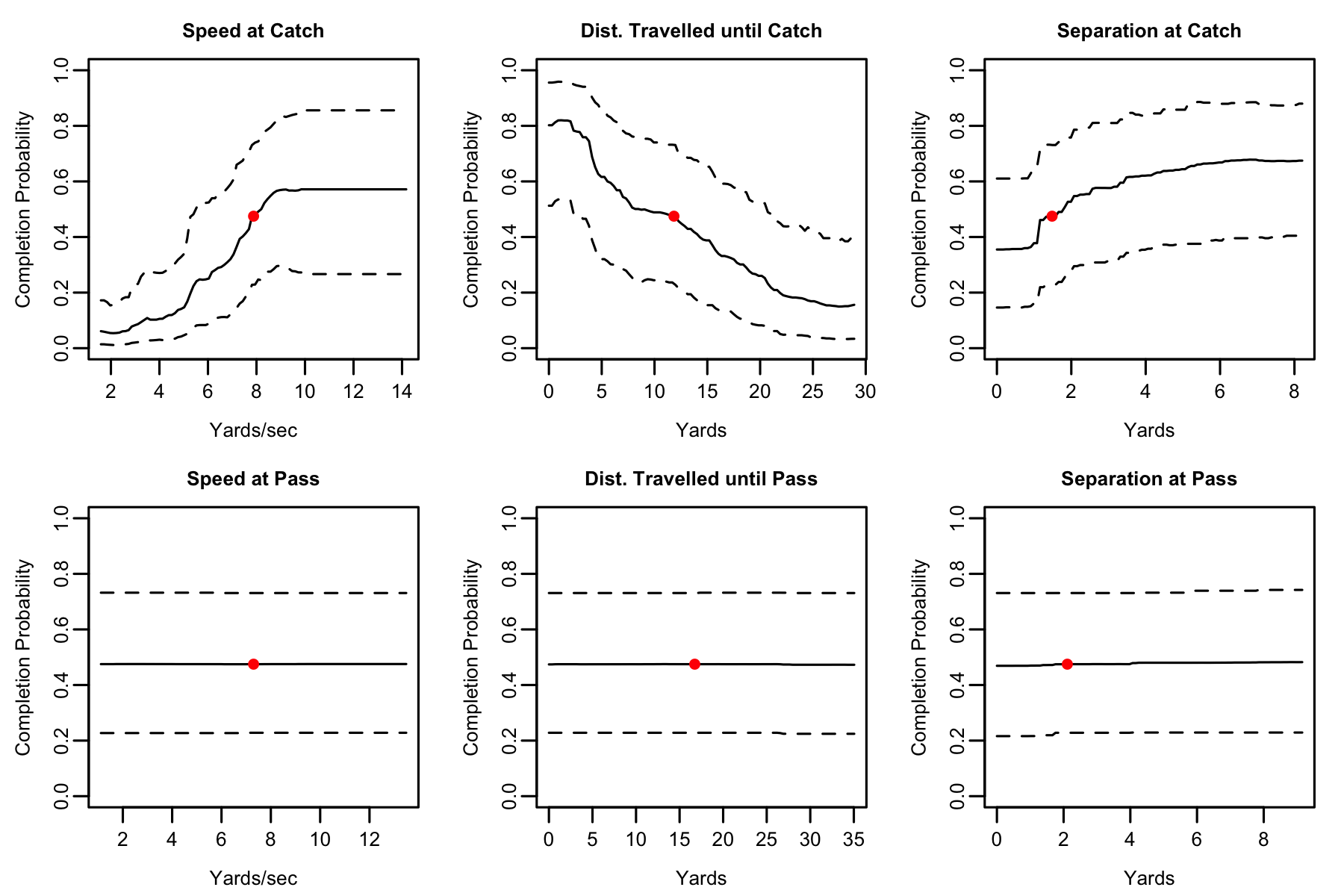}
\caption{The posterior mean completion probability on the Kupp touchdown as a function of a single covariate, keeping all other fixed. The dashed lines show the upper and lower bounds of the 95\% posterior credible interval. The actual observed value of each covariate is indicated with a red dot. Notice that the completion probability is most sensitive to variables measured at the time of the catch but not at the time of the pass.}
\label{fig:kupp_completion_prob_profiling}

\end{figure}

As suggested by an anonymous referee, another way to assess the relative importance of each of the covariates is to fix the values of all but one covariate and see how the completion probability changes as we vary that one covariate.
Figure~\ref{fig:kupp_completion_prob_profiling} illustrates how the posterior mean and 95\% credible intervals of the estimated completion probability on the Kupp touchdown mentioned earlier changes as we vary several of the covariates.
We see, for instance, that Kupp's estimated completion probability is increasing as a function of his speed at the time of the catch up until about eight yards per second, after which point the completion probability levels off.
Similarly, we see that his completion probability increases as a function of his separation at the time of the catch.
We also see that his completion probability decreases as the more he has to run between the snap and the catch.
Interestingly, however, the completion probability does not seem to be depend at all on his speed, separation, and how far he ran between the the time the ball was snapped and the pass was thrown.
This apparent lack of these three variables is further evidenced by their very low posterior splitting probabilities (0.27\%, 0.08\%, and 0.01\%).

\subsection{Simulating Unobserved Covariates}
\label{sec:simulating_x_miss}

As alluded to in Section~\ref{sec:introduction} and Figure~\ref{fig:visual_route_manuscript}, when we consider hypothetical passes, we must account for the uncertainty in the covariates that summarize what happens after the pass was thrown.
This necessity is driven home by the fact that the variables most predictive of completion probability are in fact ones measured \textit{after} the catch and are therefore not measured on hypothetical passes.
For each counterfactual pass, we first divide the covariates into two groups: those which we directly observe and those we cannot observe and about which we are uncertain.
The variables in this second group include: (i) the receiver's speed and direction at the time of the catch attempt, (ii) the pairwise Euclidean, horizontal, and vertical distances between the receiver, his nearest defender, and the ball when the catch was attempted, (iii) the total Euclidean distance the receiver travelled between the snap and the catch attempt, (iv) the total time the ball was in the air, and (v) changes in the receiver's speed, separation, direction, and total distance travelled while the ball was in the air.
Formally, let $\bx^{\star} = (\bx^{\star}_{\text{obs}}, \bx^{\star}_{\text{miss}})$ be the partition of the counterfactual covariates into the observed and missing data.
We propose to sample the values in $\bx^{\star}_{\text{miss}}$ from the empirical distribution of observed covariates. 
For instance, since we cannot observe the vector from the receiver to the ball when the hypothetical pass arrives, we randomly sample this vector from the collection of all such vectors we actually observe in the dataset. 
So if we knew the true value of $f$, the log-odds of completion function, we could approximate
\begin{equation}
\label{eq:ehcp}
\text{EHCP}(\bx^{\star}_{\text{obs}}) = \E_{\bx^{\star}_{\text{miss}}}[F(\bx^{\star}_{\text{obs}}, \bx^{\star}_{\text{miss}})] \approx \frac{1}{M}\sum_{m  = 1}^{M}{F(\bx^{\star}_{\text{obs}}, \bx^{\star(m)}_{\text{miss}})},
\end{equation}
where $\bx_{\text{miss}}^{\star(1)}, \ldots, \bx_{\text{miss}}^{\star(M)}$ are the draws of $\bx_{\text{miss}}$ from the empirical distribution, $F(\cdot) = \left[1 + e^{-f(\cdot)}\right]^{-1}$ is the forecasted completion probability function, and the expectation is taken over the empirical distribution of $\bx^{\star}_{\text{miss}}.$
Rather than setting the value of $\bx^{\star}_{\text{miss}}$ at some arbitrary fixed quantity, EHCP averages over the uncertainty in the unknown (and unobservable) values of $\bx^{\star}_{\text{miss}}.$
Importantly, since we are sampling the values of $\bx^{\star}_{\text{miss}}$ from the set of values actually observed, EHCP is constructed using realistic values of the missing covariates.

Since we do not know $f$ exactly but instead have only our MCMC samples, we can approximate EHCP for each posterior draw of $f$, thereby simulating draws from the posterior distribution of $\text{EHCP}(\bx^{\star}_{\text{obs}}).$
We can then report the posterior mean as a point estimate of the true EHCP on the hypothetical pass and also report the 95\% interval, containing likely values of the EHCP. 
We can further consider all of the routes run on a given play and track these two quantities as the play develops to see which receiver-route combinations have the highest chance of pass completion. 

\section{Illustration}
\label{sec:illustration}

To illustrate our proposed framework, we return to the two plays from the introduction, the Kupp touchdown and the Tolzien interception. 

\subsection{Completion Probability Model}

\begin{figure}[H]
\centering
\includegraphics[width = 0.45\textwidth]{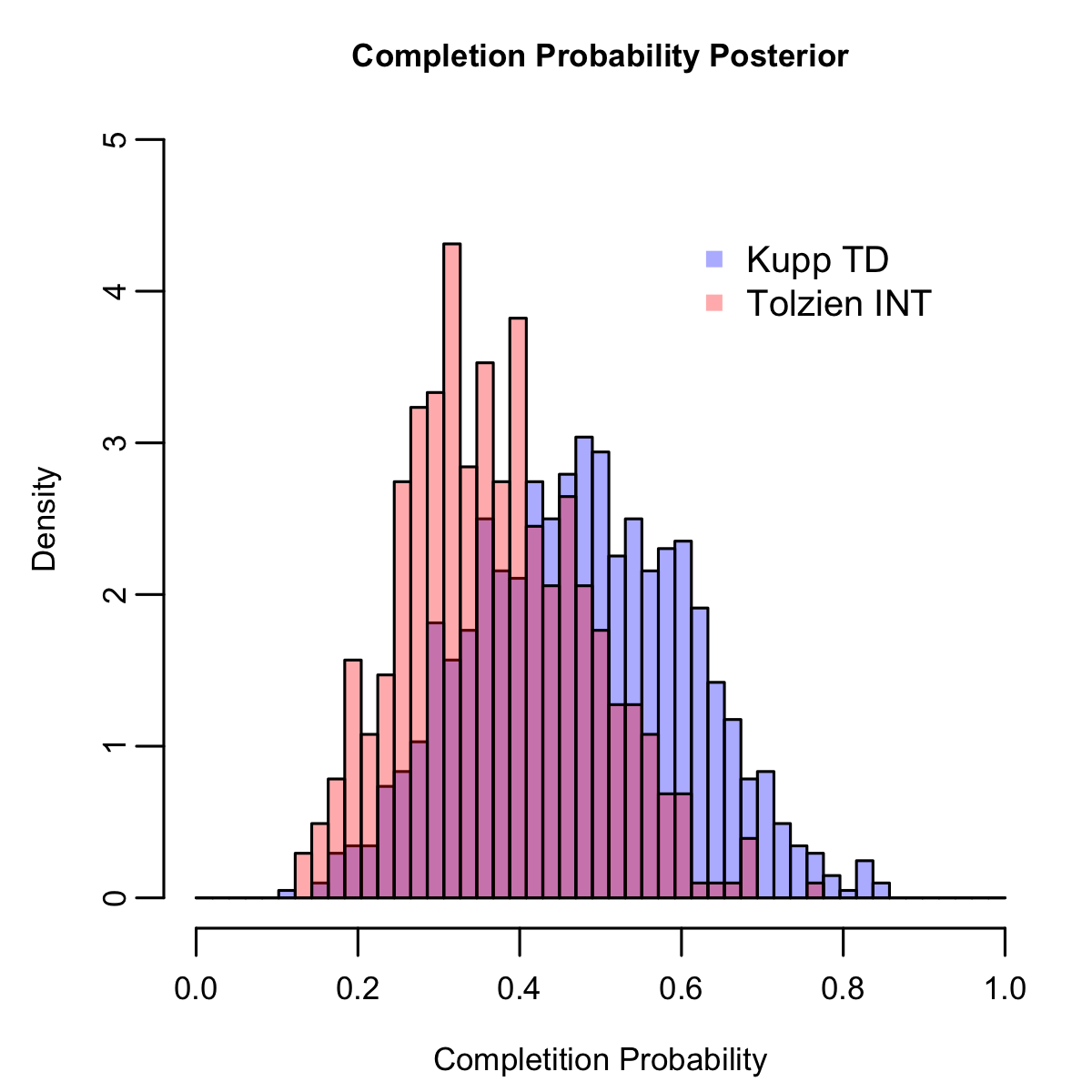}
\caption{Histogram of posterior draws of completion probabilities for the Kupp touchdown (blue) and the Tolzien interception (red)}
\label{fig:completion_posterior_hist}
\end{figure}

Figure~\ref{fig:completion_posterior_hist} shows the histogram of the posterior draws of the forecasted completion probability $F$ for the Kupp touchdown (blue) and the Tolzien interception (red).
We see that there is substantial overlap in the bulk of these posterior distributions but the posterior for the Kupp touchdown is shifted slightly to the right of posterior for the Tolzien interception.
Interestingly, on both of the these throws the receiver had less than 50\% chance of catching the ball, with the posterior mean completion probability on the Kupp touchdown approximately 10 percentage points higher than the probability for the Tolzien interception (47\% vs 37.1\%).  

\subsection{How EHCP Evolves Over A Route}

\begin{figure}[H]
\centering
\includegraphics[width = 0.45\textwidth]{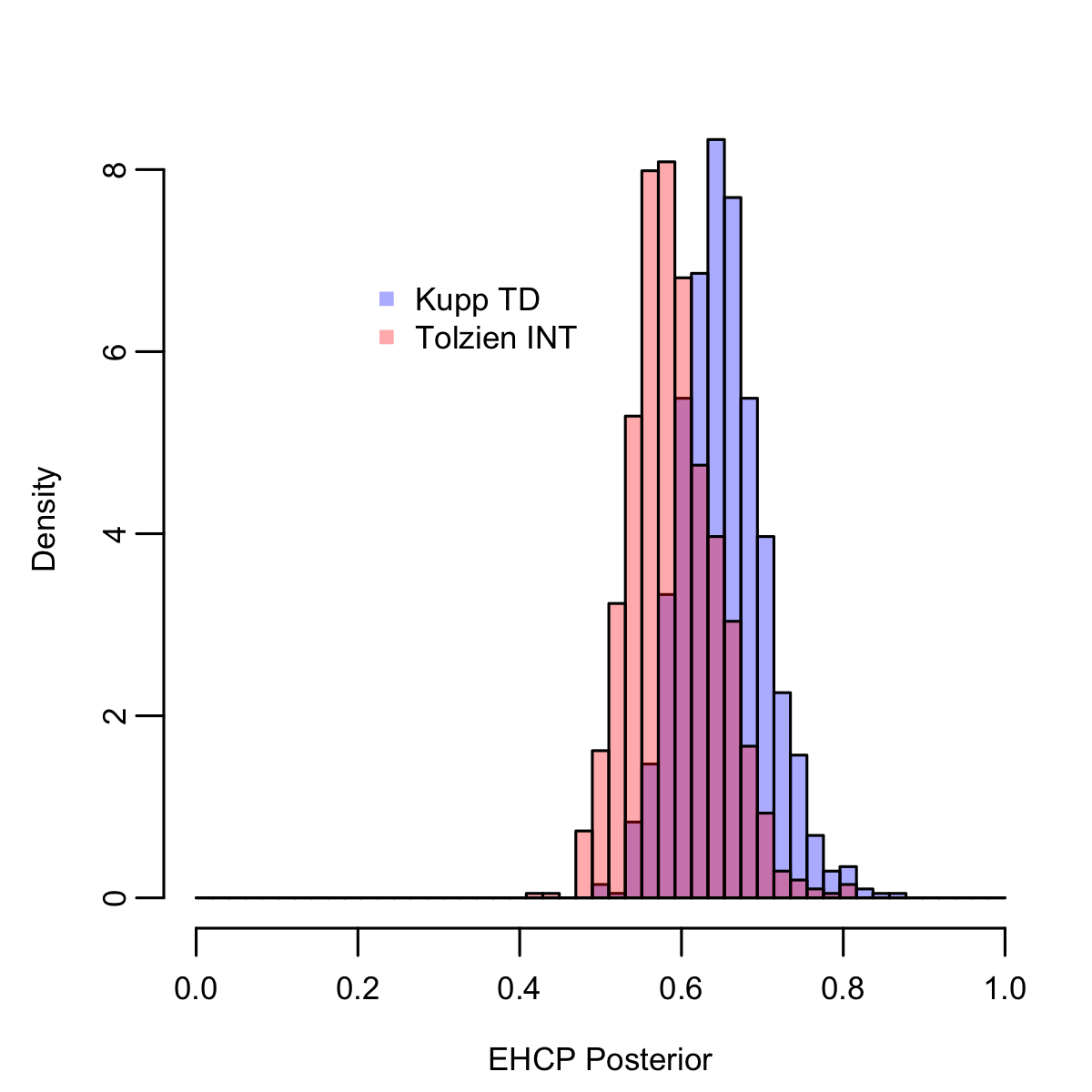}
\caption{Histogram of posterior draws of EHCP for Kupp touchdown (blue) and the Tolzien interception (red)}
\label{fig:ehcp_posterior_hist}
\end{figure}

Figure~\ref{fig:ehcp_posterior_hist} shows the histogram of the posterior EHCP draws for Kupp and Hilton (the intended target on the Tolzien interception) at the times that the two passes actually arrived. 
As before, the posterior for the Kupp touchdown is shifted slightly to the right of the Tolzien interception.
We find that the posterior mean EHCP for the Kupp touchdown is just around six percentage points higher than the posterior mean EHCP for the Tolzien interception (65.1\% vs 59.0\%).

That the EHCP and forecasted completion probabilities are somewhat different is not surprising, as they measuring two different quantities: the forecasted completion probability model uses the exact information about what actually happened after the ball was thrown while EHCP averages over the uncertainty in what might have happened after the ball was thrown.
We also note that often EHCP posteriors seem to have less variance than the posterior completion probability. 
This is also not surprising; EHCP represents an \textit{average} probability over several possible realizations of the pass while the forecasted completion probability considers only a single pass.
In a certain sense, because EHCP averages over many passes, it somewhat mitigates uncertainty introduced in our estimation of $f.$

\begin{figure}[H]
\centering
\includegraphics[width = 0.8\textwidth]{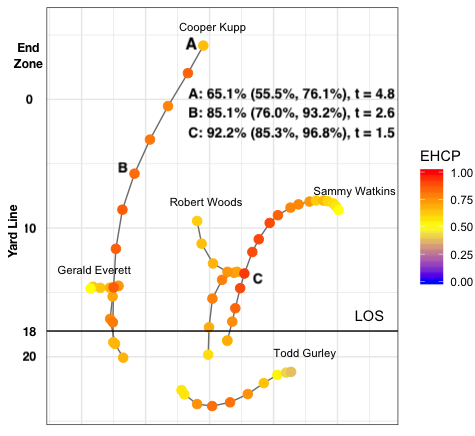}
\caption{Posterior mean EHCP for each receiver on the Kupp touchdown. 95\% posterior intervals are shown in parentheses. $t$ lists the time in seconds after the snap}
\label{fig:kupp_ehcp_trajectory}
\end{figure}

\begin{figure}[H]
\centering
\includegraphics[width = 0.8\textwidth]{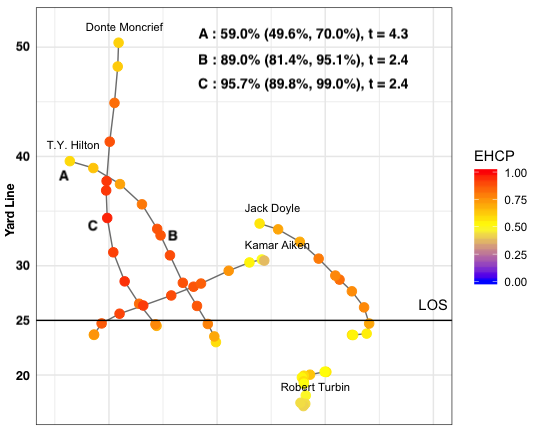}
\caption{Posterior mean EHCP for each receiver on the Tolzien touchdown. 95\% posterior intervals are shown in parentheses. $t$ lists the time in seconds after the snap}
\label{fig:tolzien_ehcp_trajectory}
\end{figure}

While comparing the EHCP for the two receivers actually targeted in the two plays at the times that the actual passes arrived is interesting, the real power of EHCP lines in projecting what might have happened had the ball been delivered to other receivers earlier in the play. 
Figures~\ref{fig:kupp_ehcp_trajectory} and~\ref{fig:tolzien_ehcp_trajectory} show the posterior mean of the EHCP for each receiver at various points in his route for the Kupp touchdown and Tolzien interception.

We see that Kupp's posterior mean EHCP at the time the actual pass arrived (location A in the figure) was 65.1\%.
Almost two seconds earlier, however, his posterior mean EHCP was 85.1\% (location B in the figure). 
Looking at the full posterior distributions of the EHCP at these two locations, we find that the 95\% intervals are nearly disjoint.
So we may conclude with reasonable certainty that Kupp's EHCP would have been higher had the pass been delivered earlier along his route.

Even more interesting, we find that of all of the receivers during this play, Sammy Watkins actually had the highest posterior mean EHCP 1.5 seconds after the snap (92.2\% at location C).
At that time, Kupp's posterior mean EHCP was 91.9\% and his 95\% interval was (85.5\%, 96.8\%), virtually identical to Watkins'.
Our analysis suggests that while the actual play resulted in a touchdown, there were times earlier in the play where the receivers would have had substantially larger expected completion probabilities. 
That being said, there are many reasons that the pass was not actually thrown to Watkins at location C. 
We will return to this point in Section~\ref{sec:discussion}.

Turning our attention to the the Tolzien interception, we find that T.Y. Hilton, the targeted receiver, had an EHCP of 59.0\% at the time the actual pass arrived (location A in the figure).
Similar to the Kupp touchdown, almost two seconds earlier, his EHCP was substantially higher (89\% at location B).
Further, Donte Moncrief had the highest EHCP of all receivers at location C, 2.4 seconds after the snap.
The substantial overlap in the 95\% intervals for Hilton ([81.4\%, 95.1\%])and Moncrief ([89.9\%, 99.0\%]) at this time means that we cannot tell with much certainty which of the two receivers had the higher EHCP.

We do note, however, that they are very close to one another on the field, which could partially explain the similarity in EHCP at that point in time. 
It is interesting to note that the posterior mean EHCPs at the time the pass actually arrived to Hilton (4.3 seconds after the snap) hovered between 40\% and 60\% for all receivers on the field.

\subsection{Player Comparisons Using EHCP}

A natural use case of EHCP is to compare players.
For example, we can examine how often a quarterback targeted the receiver with the highest or lowest EHCP on a particular play.
Such an analysis can begin to disentangle whether a declining quarterback is making bad decisions (i.e. throwing to receivers with low EHCP) or bad throws.
Table~\ref{tab:QBcomparisons} shows the results of such an EHCP-enabled quarterback analysis for the 91 games in the first six weeks of the 2017 NFL season.
In the first six weeks, Jameis Winston threw the receiver with the highest EHCP 26.8\% of the time while targeting the receiver with the lowest EHCP just 16.8\% of the time.
In contrast, Carson Wentz targeted the receiver with highest EHCP on just 15.3\% of his passes, instead throwing to the receiver with the lowest EHCP on over a quarter of all of his passes.

\begin{table}[H]
\begin{centering}
\caption{Best and worst quarterbacks at throwing to the most and least open
receivers (based on EHCP). Percentages reflect the percent of times that quarterback
threw to the most or least open receiver (required at least 100 passes).}
\label{tab:QBcomparisons}
\begin{tabular}{llrr}
\hline
 & Quarterback & Most & Least \\ \hline
\multirow{3}{*}{Best} & Jameis Winston & 26.8\% & 16.8\%\\ 
 & Trevor Siemian & 26.2\% & 16.9\% \\
 & Jared Goff & 24.8\% & 19.2\% \\
\hline 
\multirow{3}{*}{Worst} & Russell Wilson & 13.2\% & 23.5\%\tabularnewline
 & Derek Carr & 15.1\% & 27.5\%\tabularnewline
 & Carson Wentz & 15.3\% & 27.5\%\tabularnewline
\hline 
\end{tabular}
\par\end{centering}
\end{table}

Beyond looking at quarterbacks, we can also evaluate receivers using EHCP.
Compare the EHCP to the fitted completion probability from our BART model helps begin to quantify how much credit (resp. blame) the receiver ought to receiver for making (resp. failing to) make a catch.
Indeed, if a pass has a very high EHCP but very low estimated completion probability, we can infer that some combination of the receiver's actions and the defense's reactions \textit{after} the pass was thrown has reduced his chances of catching the ball.
On the other hand, if a pass has a very low EHCP but very high estimated completion probability, we can conclude that the receiver's actions and defense's reactions while the ball was in the air have improved the receiver's chances of catching the pass.
In this way, a comparison of EHCP and the estimated completion probability provides a quantitative bound on how much credit or blame to assign to the receiver's actions after the ball was thrown. 
Table~\ref{tab:RECcomparisons} shows the receivers with the highest (resp. lower) average differences between their EHCP and estimated completion probability over the six weeks in our dataset.
We find that on average Golden Tate's EHCP was about 11.8 percentage points lower than the estimated completion probability, suggesting that his actions and the defenses' corresponding reactions generally improved his chances of catching these balls. 
On the other hand, whatever DeAndre Hopkins and the defenders covering him did while the ball was in the air generally decreased his chances of completing the pass, on average.
It is important to stress, however, that the analyses contained in Tables~\ref{tab:QBcomparisons} and~\ref{tab:RECcomparisons} are for illustrative purposes only and represent only six games worth of data for each player.
As such, we do not recommend extrapolating much from these results.

\begin{table}[H]
\begin{centering}
\caption{The receivers with the highest and lowest average differences between EHCP and estimated catch probability (required at least 40 targets).}
\label{tab:RECcomparisons}
\begin{tabular}{lrrr}
\hline
Receiver & EHCP & Observed Catch Probability & Difference\tabularnewline
\hline 
Golden Tate & 64.9\% & 76.7\% & 11.8\%\tabularnewline
Christian McCaffrey & 65.1\% & 75.0\% & 9.9\%\tabularnewline
Antonio Brown & 60.2\% & 68.7\% & 8.5\%\tabularnewline
\hline 
Dez Bryant & 60.9\% & 42.5\% & -18.4\%\tabularnewline
DeAndre Hopkins & 60.5\% & 42.6\% & -17.8\%\tabularnewline
Keenan Allen & 64.8\% & 54.2\% & -10.6\%\tabularnewline
\hline 
\end{tabular}
\par\end{centering}
\end{table}

\section{Discussion}
\label{sec:discussion}

As presented here, EHCP provides an objective way to evaluate offensive plays retrospectively.
Specifically, we can track how the completion probability evolves for each receiver over the course of a play in a way that accounts for the uncertainty about missing covariates. 
The EHCP framework can also be used prospectively.
A defensive coordinator might, for instance, ask how best to cover a particular set of routes being run.
She may fix some of the unobserved covariates like the defender's position relative to targeted receiver and then average over the uncertainty in the remaining covariates to derive the EHCP for that particular combination of receiver-defender positioning. 
Repeating this for various defender locations would enable her to construct optimal defender trajectories that minimize the intended receiver's EHCP. 

Our completion probability model and the EHCP framework can also be used to provide more nuanced broadcast commentary. 
In particular, if there was a play where the forecasted completion probability and EHCP were high and the receiver failed to the catch the ball, one may reasonably assign some amount of blame to the receiver for not catching the ball; after all, the route was run and the ball was delivered to give him a high probability of catching it.
On the other hand, if the receiver catches a ball with very low forecasted completion probability and EHCP, it would be worthwhile to point out that receiver is succeeding despite the route design and pass delivery. 
Finally, one could aggregate the discrepancy between outcome and EHCP over all of a receiver's targeted routes to measure how the receiver is executing his assigned routes.

We note that the NFL's Next Gen Stats include a Completion Probability metric that is similar to our forecasted completion probability but uses different input variables than us.
Notably, Completion Probability includes a number of quarterback-centric features such as speed of and distance to the nearest pass rusher at the time of the throw \citet{NGS2018}.
Since quarterback pressure affects where the pass ends up (e.g. if it is over- or under-thrown), EHCP accounts for it rather indirectly in averaging over the uncertainty in the ball's position relative to the receiver.
That said, incorporating variables about the delivery of observed passes directly into the completion probability model is straightforward as is simulating the unobserved values of these variables for counterfactual passes in the EHCP calculation.
Doing so would result in an EHCP that better accounts for why balls were thrown when they were and would enable more nuanced assessment of the hypothetical passes. 
We hope that our method, and our transparency about how we developed it, will facilitate further iterations that combine information about the quarterback and all receivers. 

There are several potential areas of methodological and modeling improvement.
It is quite straightforward to include more covariates about the individual players involved in the pass completion model such as their historic completion percentage or how many times they have been targeted in the game so far.
We could also include more situational variables like the expected number of points of win probability estimated from models available in the \texttt{R} package ``nflscrapR'' \citep{nflscrapR}.
While we have focused on completion probability as the metric by which to assess routes, it is possible to derive analogous measures for different route metrics.
For instance, instead of modeling whether the pass was caught or not, we could model whether the play resulted in a first down and derive the expected hypothetical first down probability.
We could also model a continuous outcome like change in win probability or change in expected points scored.
Operationally, it is very straightforward to alter our code to handle continuous outcomes.
These other measures can more directly address the question of assigning expected values to particular play designs and route combinations.

More substantively, a more sophisticated imputation model of $\bx^{\star}_{\text{miss}}$ could lead to more accurate EHCP estiamtes.
In the present paper, we have taken by far the simplest approach and sampled $\bx^{\star}_{\text{miss}}$ from the observed distribution from \textit{all passes} in our dataset. 
It would be interesting to construct predictive models of $\bx^{\star}_{\text{miss}}$ using the observed covariates $\bx^{\star}_{\text{obs}}$ and to feed forecasts from these models into the EHCP calculation in Equation~\eqref{eq:ehcp}.
Doing so requires careful modeling of the joint distribution of several continuous and categorical variables, which is much beyond the current scope.

\bibliography{bdb_bib}

\end{document}